\documentclass[sn-mathphys-num]{sn-jnl}% Math and Physical Sciences Numbered Reference Style 
\usepackage{graphicx}%
\usepackage{multirow}%
\usepackage{amsmath,amssymb,amsfonts}%
\usepackage{amsthm}%
\usepackage{mathrsfs}%
\usepackage[title]{appendix}%
\usepackage{xcolor}%
\usepackage{textcomp}%
\usepackage{manyfoot}%
\usepackage{booktabs}%
\usepackage{algorithm}%
\usepackage{algorithmicx}%
\usepackage{algpseudocode}%
\usepackage{listings}%
\usepackage{xcolor}%
\usepackage{lmodern}%
\theoremstyle{plain}%
\theoremstyle{definition}  % for examples

\theoremstyle{remark}      % for remarks

\theoremstyle{definition}

\begin{document}

\title[Modeling Bias Evolution in Fashion Recommender Systems: A System Dynamics Approach]{Modeling Bias Evolution in Fashion Recommender Systems: A System Dynamics Approach}

%%=============================================================%%
%% GivenName	-> \fnm{Joergen W.}
%% Particle	-> \spfx{van der} -> surname prefix
%% FamilyName	-> \sur{Ploeg}
%% Suffix	-> \sfx{IV}
%% \author*[1,2]{\fnm{Joergen W.} \spfx{van der} \sur{Ploeg} 
%%  \sfx{IV}}\email{iauthor@gmail.com}
%%=============================================================%%

%\author[]{\fnm{Anonymous} \sur{Authors}}\email{}
 \author{\fnm{Mahsa} \sur{Goodarzi}}\email{mgoodarzi@albany.edu}

 \author{\fnm{M. Abdullah} \sur{Canbaz}}\email{mcanbaz@albany.edu}
% \equalcont{These authors contributed equally to this work.}

 \affil{\orgdiv{AI in Complex Systems Lab, Information Sciences and Technology}, \orgname{CEHC, University at Albany SUNY}, \orgaddress{
 %\street{264 ETEC 1220 Washington Avenue}, 
 \city{Albany}, \postcode{12226}, \state{NY}, \country{United States}}}

%\affil[2]{\orgdiv{Department}, \orgname{Organization}, \orgaddress{\street{Street}, \city{City}, \postcode{10587}, \state{State}, \country{Country}}}

%%==================================%%
%% Sample for unstructured abstract %%
%%==================================%%

\abstract{Bias in recommender systems not only distorts user experience but also perpetuates and amplifies existing societal stereotypes, particularly in sectors like fashion e-commerce. This study employs a dynamic modeling approach to scrutinize the mechanisms of bias activation and reinforcement within Fashion Recommender Systems (FRS). By leveraging system dynamics modeling and experimental simulations, we dissect the temporal evolution of bias and its multifaceted impacts on system performance. Our analysis reveals that inductive biases exert a more substantial influence on system outcomes than user biases, suggesting critical areas for intervention. We demonstrate that while current debiasing strategies, including data rebalancing and algorithmic regularization, are effective to an extent, they require further enhancement to comprehensively mitigate biases. This research underscores the necessity for advancing these strategies and extending system boundaries to incorporate broader contextual factors such as user demographics and item diversity, aiming to foster inclusivity and fairness in FRS. The findings advocate for a proactive approach in recommender system design to counteract bias propagation and ensure equitable user experiences.}

\keywords{Fashion Recommender Systems, Bias, Dynamic System Simulation}

%%\pacs[JEL Classification]{D8, H51}

%%\pacs[MSC Classification]{35A01, 65L10, 65L12, 65L20, 65L70}

\maketitle
\newpage
\section{Introduction}\label{sec1}

Recommender systems are a cornerstone of artificial intelligence applications, boasting a robust track record in the field. Hence, the recent shifts in research priorities have increasingly focused on enhancing algorithmic fairness and implementing bias mitigation strategies. While a substantial body of work has concentrated on optimizing accuracy and system performance, there remains a notable scarcity of studies specifically addressing the manifestations and impacts of bias within AI systems. Moreover, the exacerbation of these biases through feedback loops underscores the need for investigations into the dynamic, rather than static, nature of these systems in real-world scenarios \cite{shi_relieving_2024, paparella_reproducibility_2023}. In response to this gap, our paper explores the temporal dynamics of bias within Fashion Recommender Systems (FRS), employing system dynamics modeling and experimental simulations to understand and mitigate biases \cite{malitesta_popularity_2023, mansoury_feedback_2020}. We focus on FRS because their rapidly evolving trends, diverse user tastes, and strong visual/components expose biases more starkly than other domains.

The analysis of human beauty standards and the historical changes in fashion products and their representations in the media reveal the deep-seated biases inherent in the fashion industry. These biases, which manifest as preferences for certain body shapes, skin tones, and facial features, often lead to negative body images and misrepresent diverse identities \cite{ding-deconstructing}. Indeed, the concept of bias in AI, as it pertains to fashion, is not limited to overt discriminatory behaviors but also includes preconceived limitations of AI in recognizing diverse forms of beauty, the intricate nature of cultural identities, and various data quality issues that affect outcome accuracy and objectivity \cite{Karakaplan_GENDERBIAS, pachouly_role_2024, Siva_generative}.

On the other hand, fashion recommendation systems(FRS) often decontextualize ‘fashion identity’ by prioritizing statistical objectives over the nuanced understanding of individual user identities. This reductionist approach can inadvertently reinforce biases, ignoring the complex interplay of gender, race, body shape, and cultural norms, which are crucial for enhancing user satisfaction and shaping brand perception \cite{onitiu_algorithmic_2022}. In contrast, considering user feedback and socio-cultural contexts in the algorithmic modeling process results in more inclusive and representative recommendations \cite{deldjoo_review_2023}.

In addition, the increase in diversity of recommendations and product exposure can enhance customer engagement and help mitigate inherent social biases. However, there exists a delicate balance between maintaining system accuracy and achieving fairness. Integrating personality-based data into FRS can potentially enhance user satisfaction, albeit at the potential cost of reduced precision \cite{paryudi_performance_2022}.

Furthermore, feedback loops in FRS can amplify existing biases, such as popularity and presentation biases, through the recurrent use of biased data and interactions. By incorporating multiple types of interactions—ranging from individual item selections to entire outfit choices—and contextualizing these within user-specific data frameworks, FRS can begin to effectively counteract these biases \cite{celikik_reusable_2022-1, gervasi_stacking_2022, wu_multi-modal_2022, elsayed_end--end_2022}.

Please note that we assume that any debiasing intervention (e.g., re‐balancing, regularization) yields a net improvement in recommendation quality. This assumption guides our feedback‐loop analysis; its limitations are noted in Section 5. In this paper, we aim to make a significant contribution to the literature by addressing the temporal dynamics of biases within Fashion Recommender Systems (FRS). Through system dynamics modeling and experimental simulations, we have developed a comprehensive conceptual framework that identifies and mitigates the reinforcing effects of biases using a multidimensional feedback approach. By integrating user, item, and interaction data, our dynamic framework systematically explores and addresses potential origins of bias while adapting to evolving user behaviors and shifting market trends. Please note that we specifically chose FRS due to their unique blend of rapidly changing trends, diverse user preferences, and complex item-user interactions. However, our model is also applicable to other domains, such as music, movie, and e-commerce recommender systems, where similar dynamics impact recommendation quality and fairness. %This adaptability makes our approach especially valuable, providing a robust solution that ensures fairer and more accurate recommendations by examining the complex interactions and effects of biases and testing interventions at these critical “hot spots.”

Several recent works address dynamic biases in recommender systems. For instance, \cite{zhang_consumption_2020} proposes a probabilistic framework to disentangle popularity and position biases through counterfactual evaluation in a dynamic setting. Similarly, \cite{dean-recommender} develops a graph‐based method that tracks bias evolution over time by continuously updating user–item interactions. However, these approaches focus primarily on localized bias estimation rather than capturing the full feedback‐loop dynamics across data, user behavior, and algorithm updates. In contrast, our system‐dynamics model explicitly simulates the nonlinear interactions among data‐ and design‐induced biases, user interactions, and recommendation performance within Fashion Recommender Systems.

\section{Key Concepts: Bias and Recommendation Quality in Fashion Recommender Systems}

To effectively address bias and enhance recommendation quality in Fashion Recommender Systems (FRS), it is crucial to first clarify the key concepts that underpin this study. Bias plays a central role in shaping the performance and overall quality of recommender systems, influencing various nodes within these systems. In this section, we define and contextualize the different types of biases relevant to FRS, considering both explicit and unintentional forms. By clearly distinguishing these biases and exploring their implications on recommendation quality, we lay the foundation for understanding the dynamics of bias within FRS and justify the modeling choices and mitigation strategies discussed later in the paper.

\textbf{Bias in Recommender Systems:} Bias in recommender systems refers to systematic and unfair tendencies that influence the recommendations presented to users. In the context of Fashion Recommender Systems (FRS), biases can stem from various sources, each with unique implications. Position bias, for example, occurs when items displayed in prominent positions (e.g., at the top of a recommendation list) receive disproportionately more attention than those listed further down \cite{10.1007/978-3-319-78105-1_37}. This bias is particularly impactful in FRS, where visual placement heavily influences user interactions and purchasing decisions. Another common issue is popularity bias, where algorithms tend to favor items that are already popular, further boosting their visibility while sidelining niche or diverse styles \cite{10.1007/978-3-031-09316-6_1}. This homogenization effect limits exposure to underrepresented products, reinforcing trends that may not align with the varied tastes and identities of all users.

Biases within recommender systems can be categorized based on their origin and intent. Explicitly injected biases are deliberately introduced by enterprises to influence user behavior and achieve specific business objectives \cite{explicit_feedback}. While intentional, this type of bias can be controlled and strategically leveraged to benefit the business. On the other hand, unintentional biases arise from errors or oversights in data collection, skewed training datasets, or flaws in the algorithm’s design \cite{JIN2023101906}. Such biases negatively impact the system’s performance, fairness, and user experience, often leading to ethical concerns and the marginalization of specific user groups. Understanding these distinctions is key to developing strategies for mitigating bias while maintaining system objectives.

Importantly, these biases are not isolated but often interact in complex ways. For instance, position and popularity biases can collectively skew the visibility of items, leading to feedback loops where popular items continue to dominate \cite{10.1145/3397271.3401083}, further marginalizing diverse or less conventional fashion choices. These interactions underscore the need for a nuanced approach to bias mitigation that considers both the origins of bias and their dynamic evolution over time.

\textbf{Recommendation Quality:} Recommendation quality in FRS is generally assessed by evaluating the relevance, diversity, and personalization of suggested items. However, it is essential to consider different dimensions of quality, especially when bias mitigation is a priority. Relevance, typically linked to precision and accuracy, refers to how well the recommended items align with the user’s preferences and needs. This relevance can be compromised when biases distort what is deemed “relevant,” often due to skewed data or historical trends. Diversity, on the other hand, involves offering a broad range of recommendations that span various styles, categories, and cultural perspectives. High-quality recommendations should not only be relevant but also ensure users are exposed to a diverse array of options, which is crucial in fashion, where personal expression and cultural identity are key factors in user satisfaction \cite{10.1145/3460231.3474236}. Fairness is another critical aspect of recommendation quality, directly tied to bias mitigation. Fairness ensures that the system does not systematically disadvantage certain groups or unduly favor specific trends based solely on biased data. Achieving fairness often requires balancing competing algorithmic objectives to avoid reinforcing stereotypes or excluding underrepresented preferences \cite{10.1145/3604558}.

\textbf{Temporal Dynamics of Bias:} A core contribution of this work is the exploration of bias as a dynamic, evolving phenomenon rather than a static issue \cite{dean-recommender}. In FRS, biases can intensify over time due to feedback loops, where biased interactions lead to biased recommendations, which in turn generate further biased interactions. For example, if users consistently click on items positioned at the top of a list, position bias becomes more pronounced, encouraging the algorithm to continue prioritizing similar items. Additionally, biases in FRS are often nonlinear, meaning their effects can escalate or diminish in unpredictable ways depending on user behavior, market trends, and system interventions \cite{10.1145/3447548.3467298, shi_relieving_2024, mansoury_feedback_2020}.

\section{Methodology}\label{sec2}

Building on the understanding that bias can be both explicitly and unintentionally injected into recommender systems, our methodology focuses on modeling and controlling these biases within Fashion Recommender Systems (FRS). We acknowledge that explicit biases are often strategically introduced to align with business goals, while unintentional biases stem from systemic flaws like skewed data or inherent algorithmic limitations. Both forms of bias can significantly impact recommendation quality and fairness, leading to undesired consequences such as reinforcing stereotypes or limiting diversity in fashion recommendations.

Given the complex, nonlinear interactions between different types of bias, our approach begins by hypothesizing the dynamic and evolving nature of these biases in real-world scenarios. Specifically, we posit that biases in FRS are not static; they are subject to feedback loops and nonlinear dynamics that can either amplify or mitigate their effects depending on user behaviors and market trends. To explore this hypothesis, we develop a structured system dynamics model that represents these nonlinear interactions across various nodes of the recommender system. The model captures the propagation of bias through feedback loops and allows for simulations that reveal how different biases interact and evolve over time.

Our methodology is designed to rigorously test these dynamics by simulating various scenarios within the FRS environment. We employ experimental simulations to examine how explicit and unintentional biases emerge, interact, and evolve within the system. By simulating interventions and adjustments, we aim to understand the critical points at which bias mitigation can be most effective. Ultimately, our structured model serves as a robust tool for predicting and controlling the impact of bias in dynamic FRS settings, offering actionable insights for improving both fairness and recommendation quality.

\subsection{Nonlinear Dynamics of Bias}

In the context of system dynamics, very similar to flow charts, ‘nodes’ refer to specific points or elements within a system where data, decisions, or interactions converge. These nodes can represent users, items, feedback mechanisms, or any other critical variables that contribute to the functioning of the recommender system. When biases impact these nodes, they can propagate through the network, affecting the accuracy, fairness, and effectiveness of the recommendations provided. 

Recommender Systems (RS) experience reinforcement and amplification of biases stemming from a variety of sources. Inductive biases introduced by researchers during data collection, sampling, and processing, along with the design and modeling choices made by developers, significantly affect the distribution of training and testing datasets and ultimately, the system’s performance \cite{chen_bias_2023}. These biases in data and design are further compounded by user interactions, including selection, exposure, conformity, and position biases within recommendations \cite{zhang_consumption_2020, Wen-Entire, Jeunen-Probabilistic, chen_bias_2023}. Such interactions amplify existing biases, thereby deteriorating the quality of the recommendations.

\begin{figure}[ht]
    \centering
    \includegraphics[width=\textwidth]{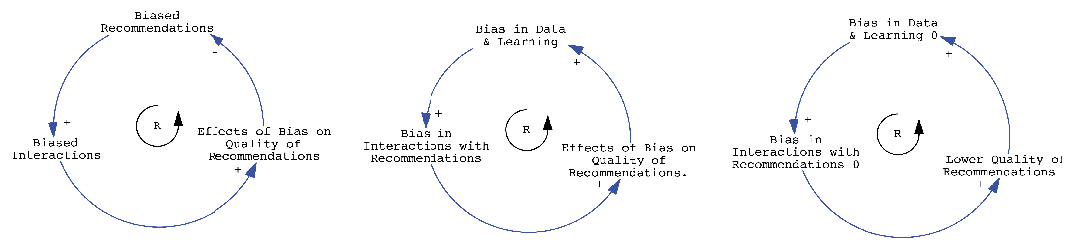}
    \caption{Feedback Loop for Reinforcing Bias Distribution}
    \label{Fig:DynamicHypothesis}
\end{figure}

Figure ~\ref{Fig:DynamicHypothesis} illustrates the direct and indirect reinforcing feedback loops in a recommender system, where biases can significantly impact the quality of recommendations. This loop consists of three main components: \textit{Biased Recommendations}, \textit{Effects of Bias on Quality of Recommendations}, and \textit{Biased Interactions}.
\begin{enumerate}
    \item \textbf{Biased Recommendations:} When a recommender system produces biased recommendations, these recommendations inherently favor certain items or preferences over others. Bias can be explicitly injected into the system as a strategic tool by enterprises to steer user behavior towards specific outcomes, such as promoting certain products, increasing sales, or enhancing user engagement.
    \item \textbf{Effects of Bias on Quality of Recommendations:} The biased recommendations subsequently affect the overall quality of the recommendations provided by the system. If the recommendations align with the intended strategic goals of the enterprise, the perceived quality might improve from a business perspective. However, if these biases are unintentional, they degrade the quality of recommendations by misrepresenting user preferences and reducing the system’s effectiveness in personalizing content.
    \item \textbf{Biased Interactions:} As users interact with the biased recommendations, their interactions become biased as well. Users might engage more with the promoted items, further reinforcing the initial bias. This biased interaction data is then fed back into the system, influencing future recommendations and perpetuating the cycle.
\end{enumerate}

Additionally, \textbf{Bias in Data and Learning}, as illustrated in Figure ~\ref{Fig:DynamicHypothesis}, encompasses the inherent prejudices and skewed distributions that are present during the data collection, processing, and model training phases. Specifically, biases in data and learning processes can result from sampling biases, where certain user groups or item categories are over- or under-represented, and algorithmic biases, where the model inherently favors certain patterns due to its training parameters and objectives. These biases propagate through the system, leading to a lower quality of recommendations. Poor quality recommendations, in turn, influence user interactions in a biased manner, perpetuating the cycle. This foundational bias has a more profound and insidious impact compared to the more observable bias in user interactions, as it sets the stage for systemic issues that compromise the entire recommendation pipeline from the outset.

\begin{figure*}[!hb]
\centering
\includegraphics[width=\textwidth]{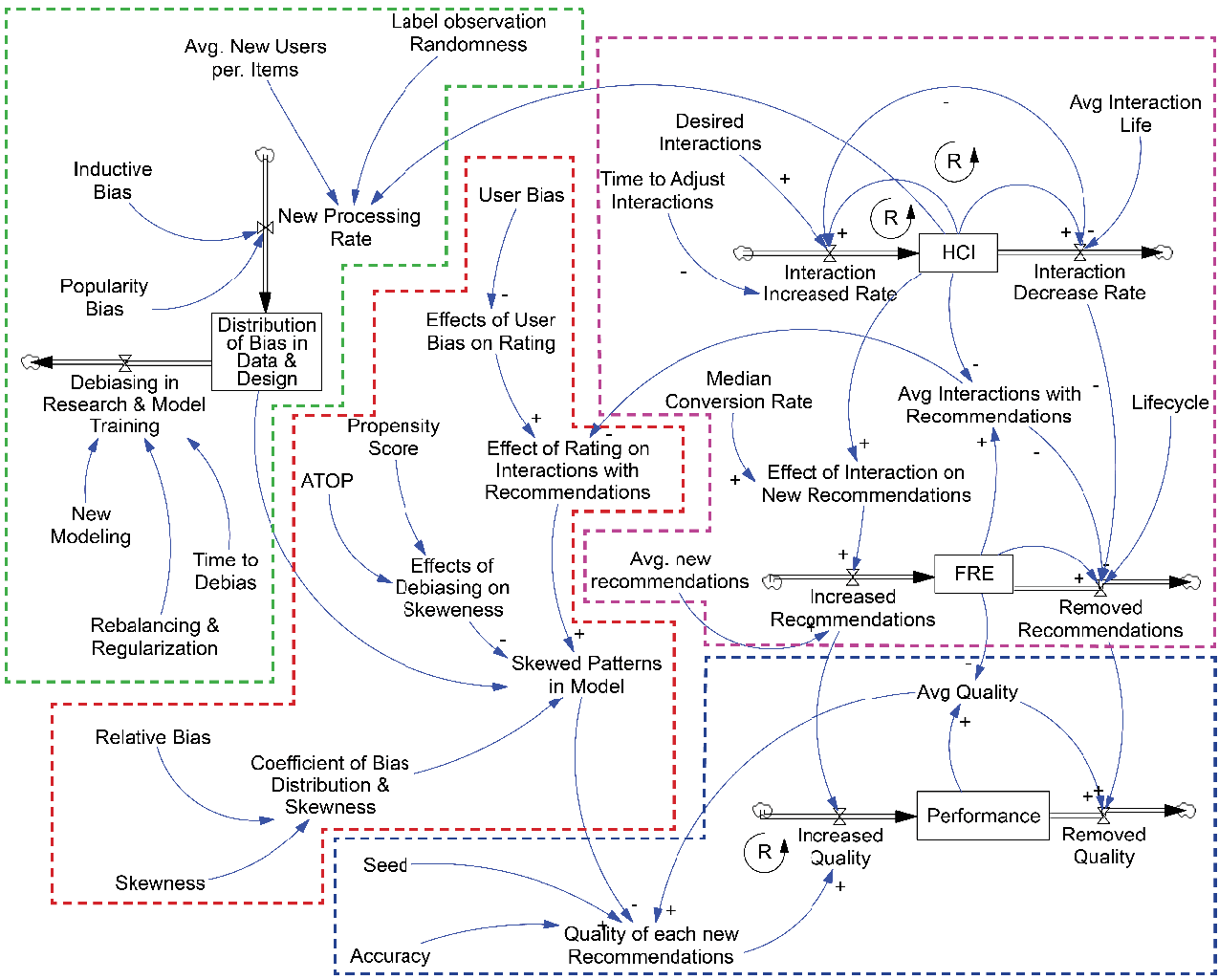}
\caption{The Stock and Flow Diagram of Bias in FRS}\label{Fig:StockFlow}
\end{figure*}

\subsection{Model Architecture}

To provide a focused and precise analysis, we developed a structured model that captures the nonlinearity of bias and allows for simulations to assess and control its impact within dynamic recommender systems. The scope of our model is constrained to internal factors that are directly involved in bias activation. These factors encompass essential processes across various components of AI systems, such as Data Management, System Design and Learning Algorithms, and Human-Interaction Dynamics.

Our work builds upon foundational methodologies presented by previous studies on dynamic recommender systems \cite{dean-recommender, Rana-dynamic, zhang_consumption_2020}. We seek to distill the complexity of bias in Fashion Recommender Systems (FRS) into abstract, manageable components through system dynamics modeling \cite{Sterman-Business}. While a more comprehensive model might include numerous biases and exogenous variables, this would have led to excessive complexity. Instead, we prioritized essential variables to establish a conceptual framework that can be expanded in future research. Moreover, we will check structural validity (e.g., sign and causal polarity of each loop) and later calibrate key coefficients using real FRS clickstream logs (e.g., from a partner retailer). Until then, the model remains conceptual.

It is important to note that we deliberately excluded exogenous factors, such as broader societal biases that often influence decision-making. While these external elements are significant in introducing bias into FRS, our study focuses on biases that emerge directly within the AI system’s decision-making processes. Grounded in existing literature, this approach allows us to concentrate on the internal mechanisms most relevant to bias dynamics \cite{Almakinah-bias}. For dynamic modeling and simulations, we utilized VensimPLE \cite{vensim-site}.

The architectural model shown in Figure ~\ref{Fig:StockFlow} illustrates these key variables and their relationships, providing a foundation for understanding and addressing bias within recommender systems. The figure serves as a visual summary of the conceptual framework that guides our analysis and simulations.

Figure ~\ref{Fig:StockFlow} highlights distinct colored boxes representing different sections of the bias dynamics model. These boxes, marked in green, red, pink, and blue, indicate the primary areas where bias originates and propagates. The green box focuses on how bias is introduced during data management and design processes, while the red box captures user behavior and how it influences skewness within the system. The pink box represents the dynamics of human-computer interaction, detailing how user engagement reinforces or mitigates bias over time. Finally, the blue box examines how bias impacts recommendation performance and quality, closing the loop by showing how biased recommendations feed back into future interactions.

These boxes are marked to allow for a clear separation of different subsystems within the overall model. By isolating these critical areas, we can better understand how bias interacts across data, design, user behavior, and system performance. Each box contributes to the overall problem by emphasizing the distinct yet interconnected mechanisms that drive bias within Fashion Recommender Systems, enabling a more targeted approach to mitigation strategies.

\subsubsection{Data Management and Design Biases (\textcolor{green}{Green Box}):}
This Green box captures the biases introduced during data management and system design stages. It includes factors like popularity bias and inductive bias that emerge as new data is processed and integrated into the system. Debiasing methods such as re-balancing and regularization are also represented here, as they aim to counterbalance these initial biases. The design and data handling processes here set the foundation for bias activation because the choices made during data selection, feature engineering, and algorithmic adjustments can either amplify existing biases or introduce new ones. This box is crucial because it represents the point of origin for biases that later cascade throughout the system.

\textbf{Contribution to Bias:} The green box is where biases are initially encoded into the system. Design choices and data curation decisions can inadvertently embed and perpetuate biases, which are then inherited by other parts of the system, leading to skewed recommendations from the outset. 
\subsubsection{User Behavior and Interaction Biases (\textcolor{red}{Red Box}):}
The red box focuses on user-related biases and the resulting skewness in recommendations. It models how user preferences, ratings, and interactions influence the overall skewness and accuracy of the recommendations. Elements like propensity scores and debiasing efforts are included here to showcase how these interventions aim to correct the bias. The box emphasizes how user behavior directly feeds back into the recommendation loop, either amplifying or mitigating biases over time. For instance, biased user ratings can distort the perceived quality of items, which in turn biases the model’s future recommendations, locking users into a feedback loop of biased content.

\textbf{Contribution to Bias:} The red box serves as the engine for bias propagation. User behaviors, influenced by initial biases, can create self-reinforcing loops that drive the system to increasingly favor certain patterns over others. This cycle gradually amplifies the skewness within the system and leads to persistent bias accumulation. We treat user biases, such as those influenced by reading reviews, as external to our model because accounting for them would require adding a separate component to explain this behavior. For our model, we consider those as constant variables with the potential to measure outcomes by changing them.

\subsubsection{Human-Computer Interaction Dynamics (\textcolor{pink}{Pink Box}):}
This Pink box represents the dynamics of how users interact with the system and how those interactions impact recommendation quality. It covers the lifecycle of interactions, the conversion rates, and the effect of interaction frequency on future recommendations. The loops in this section highlight how user engagement can reinforce bias, especially when system recommendations are shaped by biased interaction patterns. For example, users interacting more frequently with certain types of biased recommendations will lead to more similar content being presented to them, deepening the bias in subsequent iterations.

\textbf{Contribution to Bias:} The pink box plays a critical role in bias amplification through interaction loops. As user interactions become increasingly shaped by biased recommendations, the system continually adapts to serve content that reinforces those biases. This cycle escalates the bias, entrenching it further with each interaction.
\subsubsection{Performance and Recommendation Quality (\textcolor{blue}{Blue Box}):}
The blue box closes the feedback loop by focusing on how bias affects the overall performance and quality of the recommendations. It explores the impact of skewed patterns on the accuracy and relevance of recommendations, ultimately influencing user satisfaction. The feedback within this box illustrates how biased recommendations lead to more biased interactions, perpetuating the cycle. As biases degrade the quality of recommendations, user satisfaction might decline, further complicating the efforts to mitigate bias without negatively affecting system performance.

\textbf{Contribution to Bias:} The blue box is where the cumulative effects of biases become evident in terms of degraded recommendation quality and performance. The biases that flow into this section can cause the system to increasingly favor content that aligns with existing skewed patterns, making it harder to break out of biased loops. Ultimately, this box encapsulates how bias becomes a systemic issue that impacts both user experience and model accuracy.

As summary, our model shown in Figure ~\ref{Fig:StockFlow} provides a comprehensive view of how biases are initiated, amplified, and mitigated within Fashion Recommender Systems. By segmenting the model into distinct yet interconnected boxes, we capture the core dynamics that drive bias propagation—from data management and user behavior to system performance and interaction loops. The relationships depicted in the model highlight the non-linear and compounding effects of bias, offering a roadmap for identifying strategic intervention points. Through this approach, our work not only reveals the complexities of bias within FRS but also provides a foundation for enhancing fairness, accuracy, and overall system robustness in future research and practical applications. For clarity and further exploration, larger versions of these boxes are included in the appendix as individual figures.

\section{Experimental Results}\label{sec3}

In this section, we present the experimental results of our dynamic model, which demonstrates robust behavior and dimensional consistency under various initial conditions. Utilizing VensimPLE \cite{vensim-site} for dynamic modeling and simulations, we conducted sensitivity analyses to validate our findings under dynamic conditions. The results highlight the influence of bias on system performance and explore the effectiveness of different debiasing interventions.

\textbf{Experiment 1: Base Run}
In the base run, we kept all bias variables constant at a value of 1, assuming no new biases were introduced into the system. This setup allowed us to observe the model’s pure dynamics and analyze changes in quality based on initial conditions and different seed values.

\begin{figure}[!h]
    \centering
    \includegraphics[width=\textwidth]{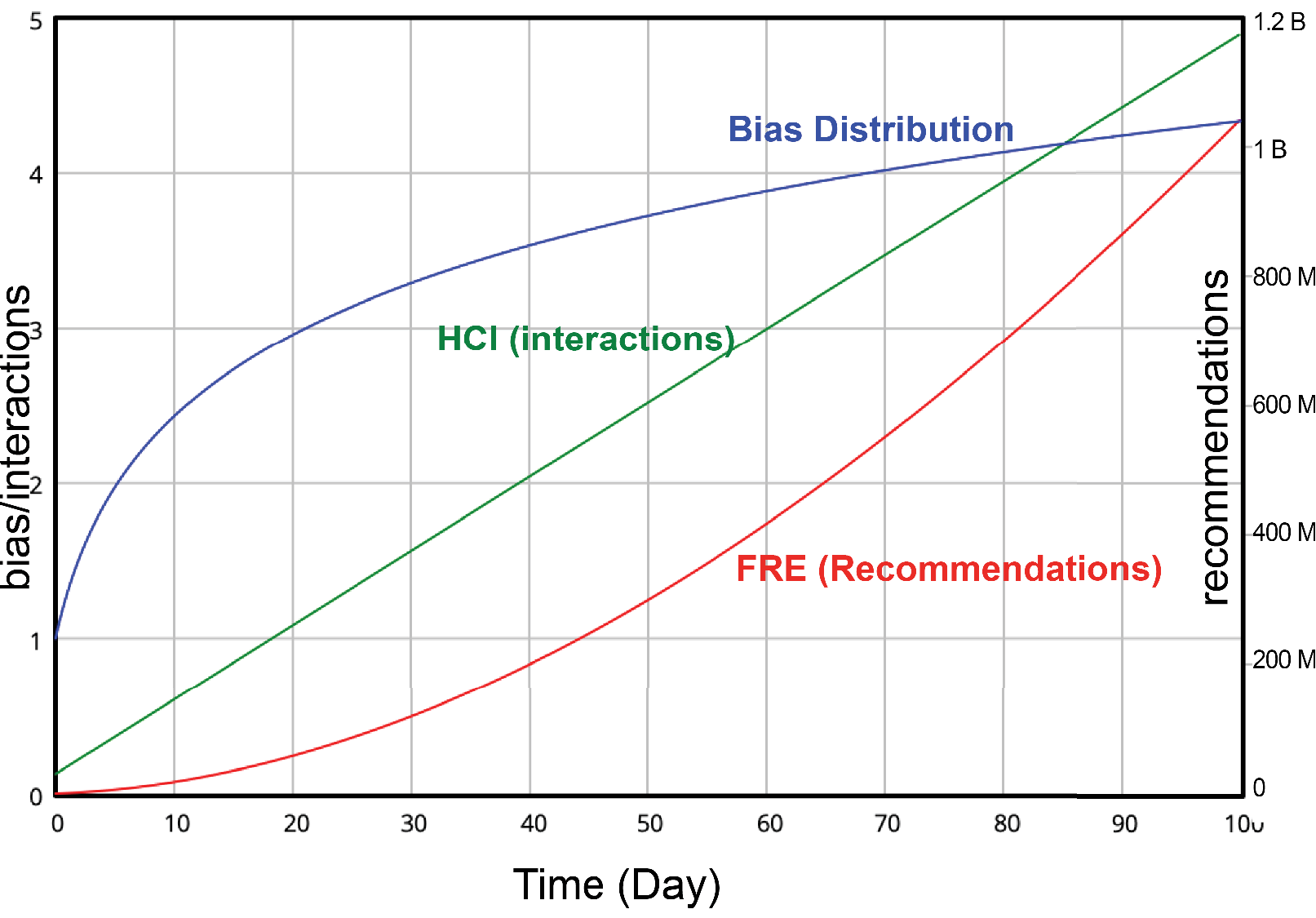}
    \caption{Base-run simulation: bias distribution, FRE recommendations, and HCI metrics over time.}
    \label{fig:experiments-baserun}
\end{figure}

As shown in Figure ~\ref{fig:experiments-baserun}, the results indicated that bias distribution exhibited rapid exponential growth initially, which then gradually slowed down over time. Meanwhile, the Fashion Recommender Engine (FRE) showed a steady, continuous increase in the number of recommendations, reflecting consistent growth throughout the simulation period. Human-Computer Interaction (HCI) metrics, representing user interactions, also demonstrated a stable upward trend, although at a different rate compared to FRE and bias distribution.

These patterns suggest that the system’s internal dynamics, particularly the interactions between bias, recommendations, and user engagements, can lead to significant variations in overall performance. Despite changes in seed values, the behavior of other variables remained stable, indicating the robustness of the model under different conditions. The observed exponential growth and eventual stabilization of bias highlight the critical need for addressing bias early in the system’s operation to maintain high-quality performance over time.

\begin{figure}[h]
    \centering
    \includegraphics[width=\textwidth]{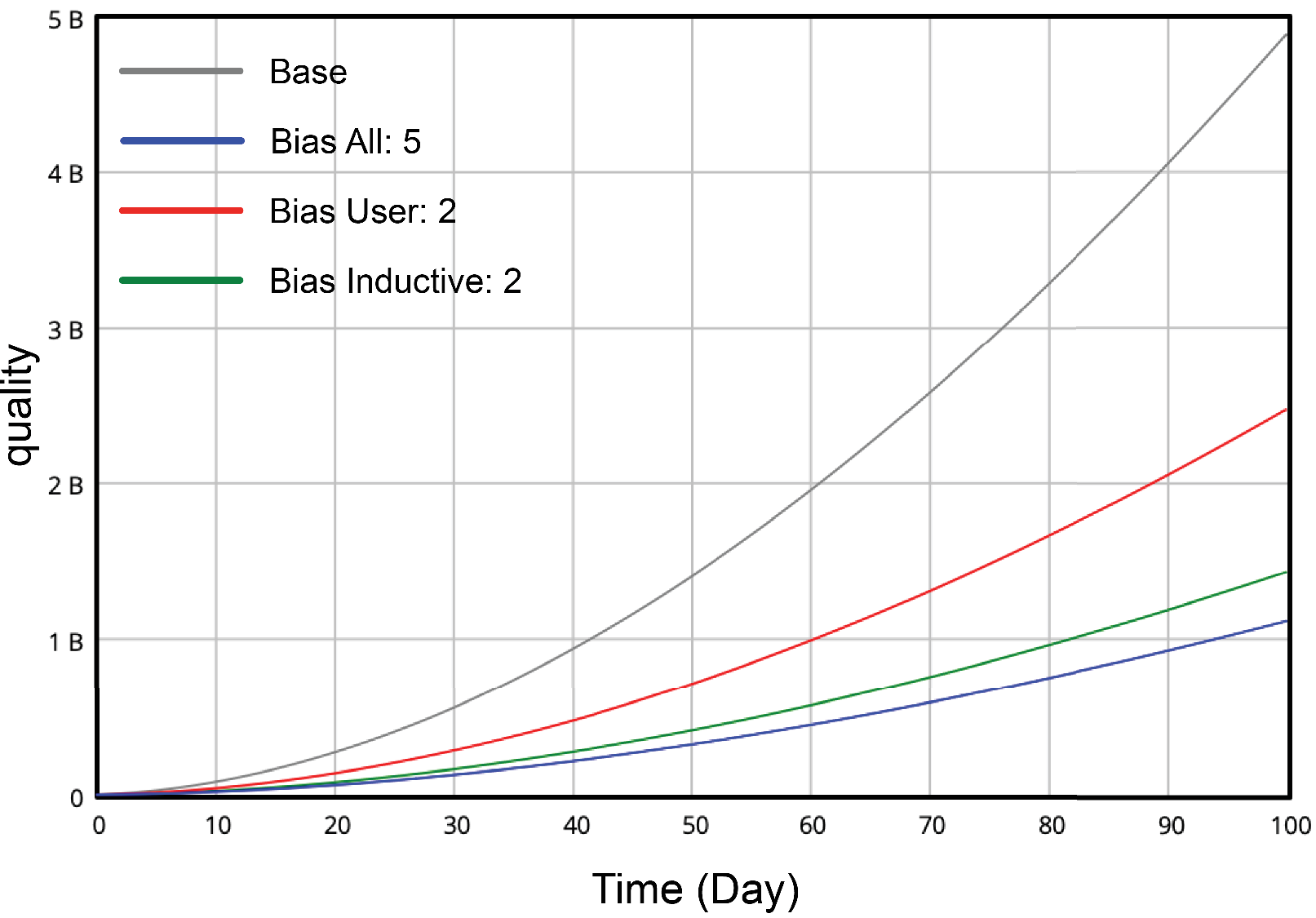}
    \caption{Impact of Bias Activation on Performance}
    \label{fig:experiments-BiasActivation}
\end{figure}

\textbf{Experiment 2: Bias Activation}
Our hypothesis suggests that introducing initial bias into the model will lead to a decrease in performance quality. Quality is the unit of performance that all system variables, whether directly or indirectly, operationalize in the context of this study. The equations leading to performance and quality count for the randomness in the system based on skewness, accuracy, and the normal distributions and variance in ratings. To test this, we conducted sensitivity analyses with different seed values, doubling the initial values of inductive and user biases in separate runs. As shown in Figure ~\ref{fig:experiments-BiasActivation}, the results indicated that the impact of increased inductive bias (initial value multiplied by 2 shown with the green line) on performance was more pronounced than that of user bias (initial value multiplied by 2 shown with the red line), primarily due to bias amplification from interactions being treated as new data.

To further investigate the model’s performance under high levels of bias, we activated all biases in one run with initial values set to five times those of the base simulation. Please note that the gray line indicates the base run, which has no biases activated. This configuration resulted in the highest bias distribution and the lowest system quality observed (shown with the blue line). However, the smaller decline in quality with high bias activation was attributed to the slower growth rate of bias distribution.

\begin{figure}[h]
     \centering
     \includegraphics[width=\textwidth]{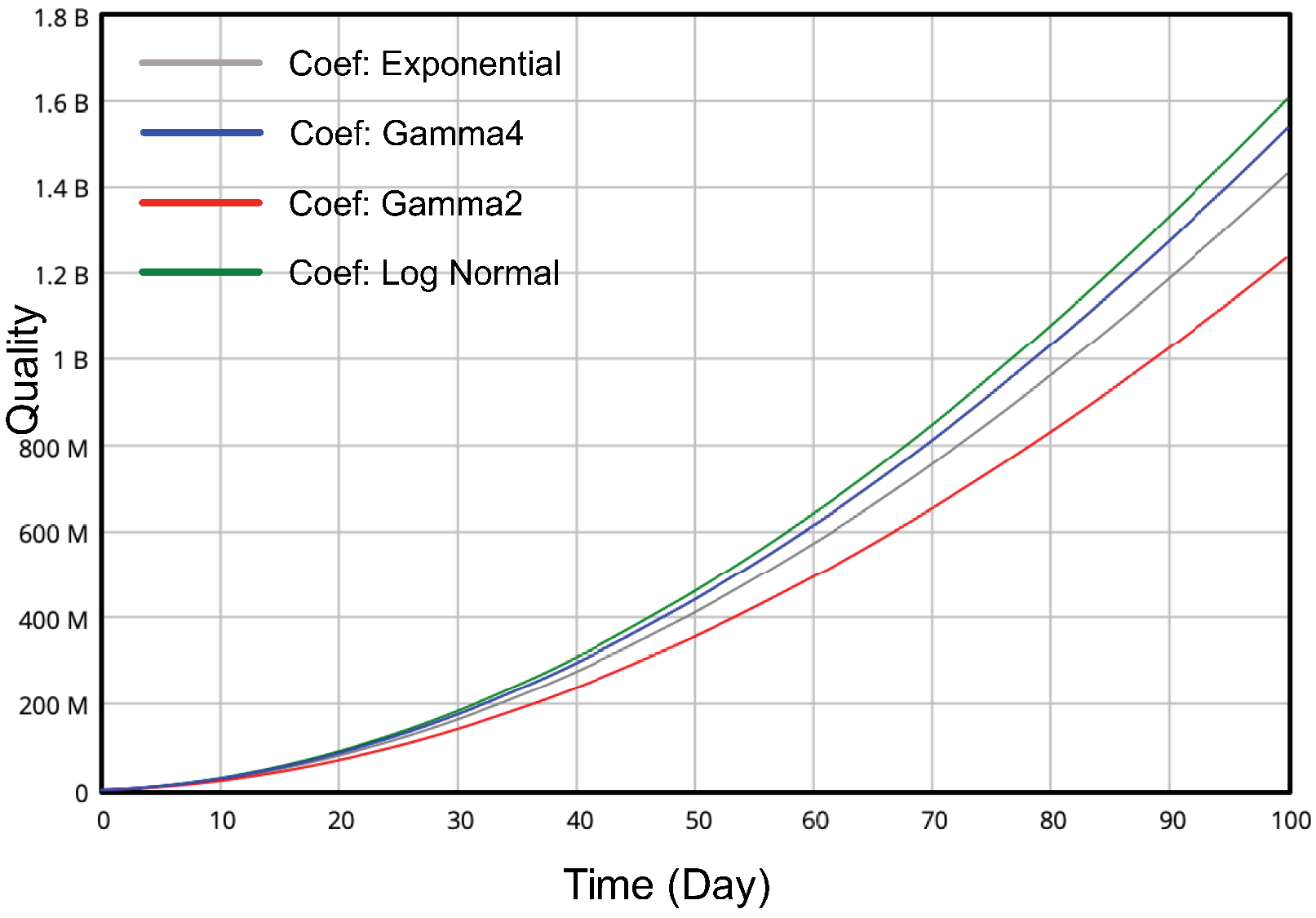}
     \caption{Impact of different coefficients for the relationship between skewness and bias distribution on quality}
     \label{Fig:Stats.png}
 \end{figure}

\textbf{Experiment 3: Statistical Distributions}
It stands to reason that with the increase in the skewness of data, the bias distribution in the system also increases. With this assumption in mind, we modeled performance under different data distribution scenarios, according to \cite{Bono-Bias}. The log-normal distribution recorded the highest average quality. The gamma distribution exhibited an interesting behavior in that its quality relied on the shape parameter $\alpha$. Lower values of $\alpha$ corresponded to lower average quality (Table ~\ref{tab:coef} \& Figure ~\ref{Fig:Stats.png}).

 \begin{table}[h]
\centering
   \caption{Skewness Per Bias Distribution}
   \label{tab:coef}
   \begin{tabular}{ccl}
     \toprule
     Distribution Type&Skewness&Relative Bias\\
     \midrule
     Exponential & 4.57 & 0.07\\
     Log Normal & 2.81& 0.14\\
     Gamma with $\alpha$ = 2 & 2.81 & 0.19\\
     Gamma with $\alpha$ = 4 & 2.04& 0.21\\
   \bottomrule
 \end{tabular}
 \end{table}

\begin{figure}[h]
    \centering
    \includegraphics[width=\linewidth]{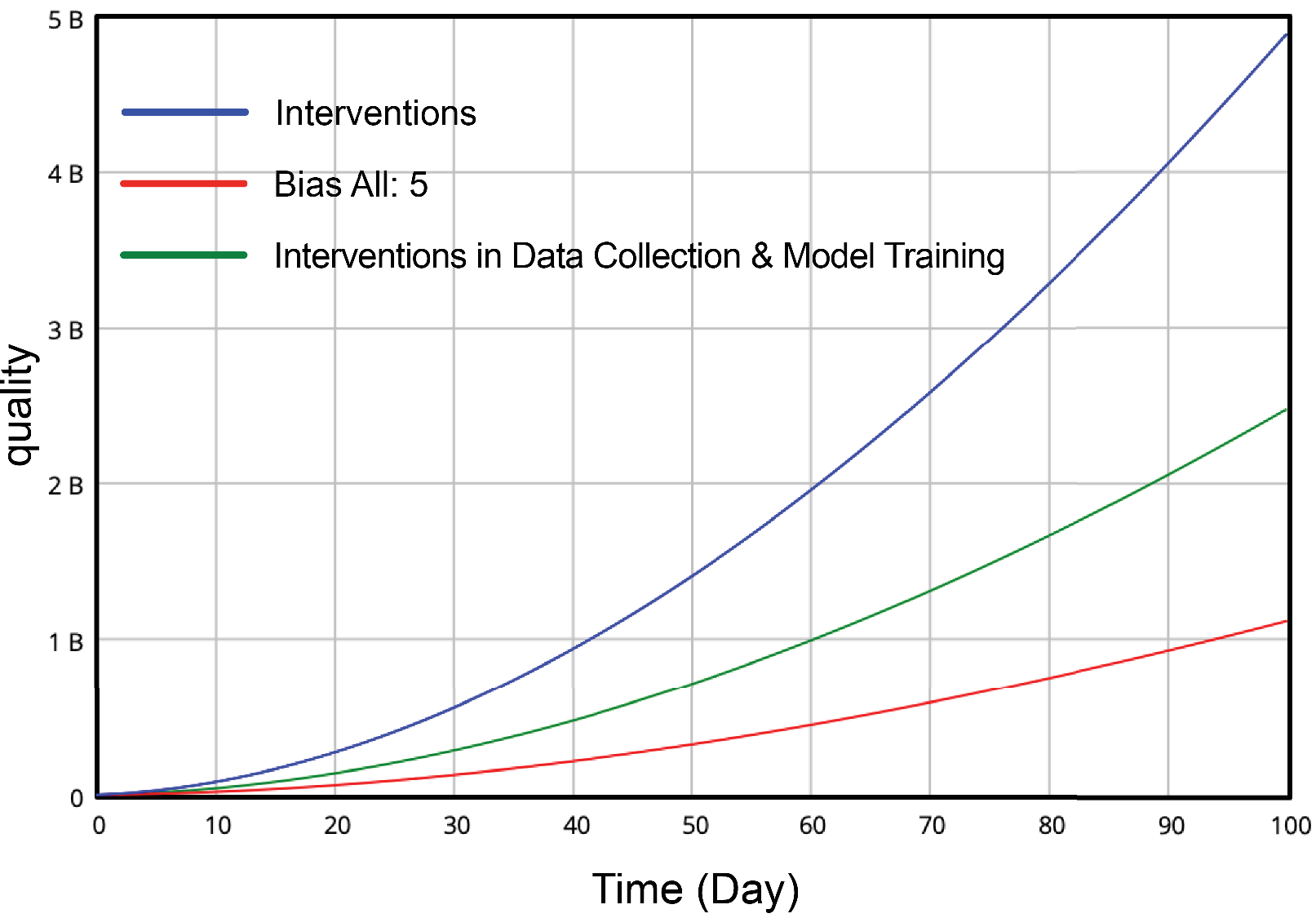}
    \caption{Impact of Interventions on Performance}
    \label{fig:experiments-BiasIntervention}
\end{figure}

\textbf{Experiment 4: Debiasing Interventions}
The significant impact of interventions on performance is illustrated in Figure ~\ref{fig:experiments-BiasIntervention}. We simulated the model with the highest level of bias, as mentioned in the previous section, and applied interventions at the same magnitude as the bias activation (represented by the blue line). These interventions included methods such as ATOP (Automatic Targeted Online Propensity) scores for user biases, re-balancing techniques, regularization, and new modeling choices in algorithmic design.

As shown in the Figure ~\ref{fig:experiments-BiasIntervention}, the blue line indicates the scenario where interventions are applied, leading to a significant improvement in performance quality over time compared to the high bias scenario (red line). The green line represents interventions specifically in data collection and model training by engineers, which also show improvement but to a lesser extent than the comprehensive interventions scenario.

%\begin{figure}[tb]
%    \centering
%    \begin{subfigure}{\textwidth}
%        \centering
%        \includegraphics[height=5.7cm]{Figures/BaseRun-1.png}
%        \caption{Base Run}
%        \label{fig:experiments-baserun}
%    \end{subfigure}
%    ~ 
%    \begin{subfigure}{\textwidth}
%        \centering
%%        \includegraphics[height=5.7cm]{Figures/Activation-1.png}
%        \caption{Impact of Bias Activation on Performance}
%        \label{fig:experiments-BiasActivation}
%    \end{subfigure}
%    ~ 
%    \begin{subfigure}{\textwidth}
%        \centering
%        \includegraphics[height=5.7cm]{Figures/Interventions-1.png}
%        \caption{Impact of Interventions on Performance}
%        \label{fig:experiments-BiasIntervention}
%    \end{subfigure}
%    \caption{Experimental Results of the Dynamic Model} 
%\end{figure}

These results suggest that implementing targeted interventions can effectively mitigate the negative impacts of bias on system performance. However, the figure also highlights the need for further research into addressing inductive biases originating from researchers, as these biases still present challenges that are not fully resolved by the current interventions.

% \begin{figure}[H]
%     \centering
%     \includegraphics[width=\columnwidth]{Figures/Interventions.png}
%     \caption{Effect of Debiasing Interventions on Quality}
%     \Description{Effect of Debiasing Interventions on Quality}
%     \label{Fig:Interventions.png
% }
% \end{figure}

\section{Discussion and Final Remarks}\label{sec4}

Despite extensive research into biases within recommender systems, the dynamic nature of these biases and their long-term impacts on system functionality remain underexplored. In this study, we use Fashion Recommender Systems (FRS) as a specific use case to explore the underlying structures of bias reinforcement and evaluate their influence on system performance.

The dynamic model of bias activation in recommender systems reveals the underlying structures of bias reinforcement and its impact on system performance. This model highlights the path dependencies in bias behavior within AI systems, showing how biases can be self-reinforcing over time. Our findings suggest that inductive biases have a more pronounced and critical effect on performance compared to user biases. Additionally, existing interventions aimed at mitigating these biases show significant potential for balancing the model over time.

However, it is important to acknowledge that the proposed model is a simplification of reality. To better control biases, we propose a forward feedback strategy that implements interventions by extending model boundaries to include additional factors such as accuracy, sales, users, and items. In the context of FRS, other potential interventions worth exploring include utilizing influencers during the adoption process and diversifying item or feature selections.

Our study emphasizes the need for further research into the relationship between bias distribution and skewness in recommender systems. By employing system dynamic modeling and experimental simulations, we observed the nonlinear behavior of bias distribution over time and its consequential impact on performance quality. The model clearly demonstrates the prominent influence of inductive biases over user biases and underscores the effectiveness of mitigation strategies in controlling the effects on system quality, even if they do not directly address bias distribution.

\newpage

\section{Limitations}\label{sec5}
This research is a work in progress that bridges several disciplines and calls for collaboration among professionals from different areas of expertise. While tested for structural and behavioral validity, the presented model requires further expansion and demands testing under extreme conditions and consideration of policy implications. It needs to extend from a conceptual system dynamics framework to support applications and policy development in complex systems, such as AI.

%\bmhead{Supplementary information}

%Not applicable

%\bmhead{Acknowledgements}

%Not applicable

%\section*{Declarations}

%All authors certify that they have no affiliations with or involvement in any organization or entity with any financial interest or non-financial interest in the subject matter or materials discussed in this manuscript.

\newpage
\bibliography{sn-bibliography}% common bib file
%% if required, the content of .bbl file can be included here once bbl is generated
%%\input sn-article.bbl

%%===================================================%%
%% For presentation purpose, we have included        %%
%% \bigskip command. Please ignore this.             %%
%%===================================================%%
\bigskip
\begin{flushleft}%
Editorial Policies for:

\bigskip\noindent
Springer journals and proceedings: \url{https://www.springer.com/gp/editorial-policies}

\bigskip\noindent
Nature Portfolio journals: \url{https://www.nature.com/nature-research/editorial-policies}

\bigskip\noindent
\textit{Scientific Reports}: \url{https://www.nature.com/srep/journal-policies/editorial-policies}

\bigskip\noindent
BMC journals: \url{https://www.biomedcentral.com/getpublished/editorial-policies}
\end{flushleft}

\newpage

\begin{appendices}

\section{Closer Look at the Stock and Flow Diagram}\label{secA1}

\begin{figure}[h]
    \centering
    \includegraphics[width=\textwidth]{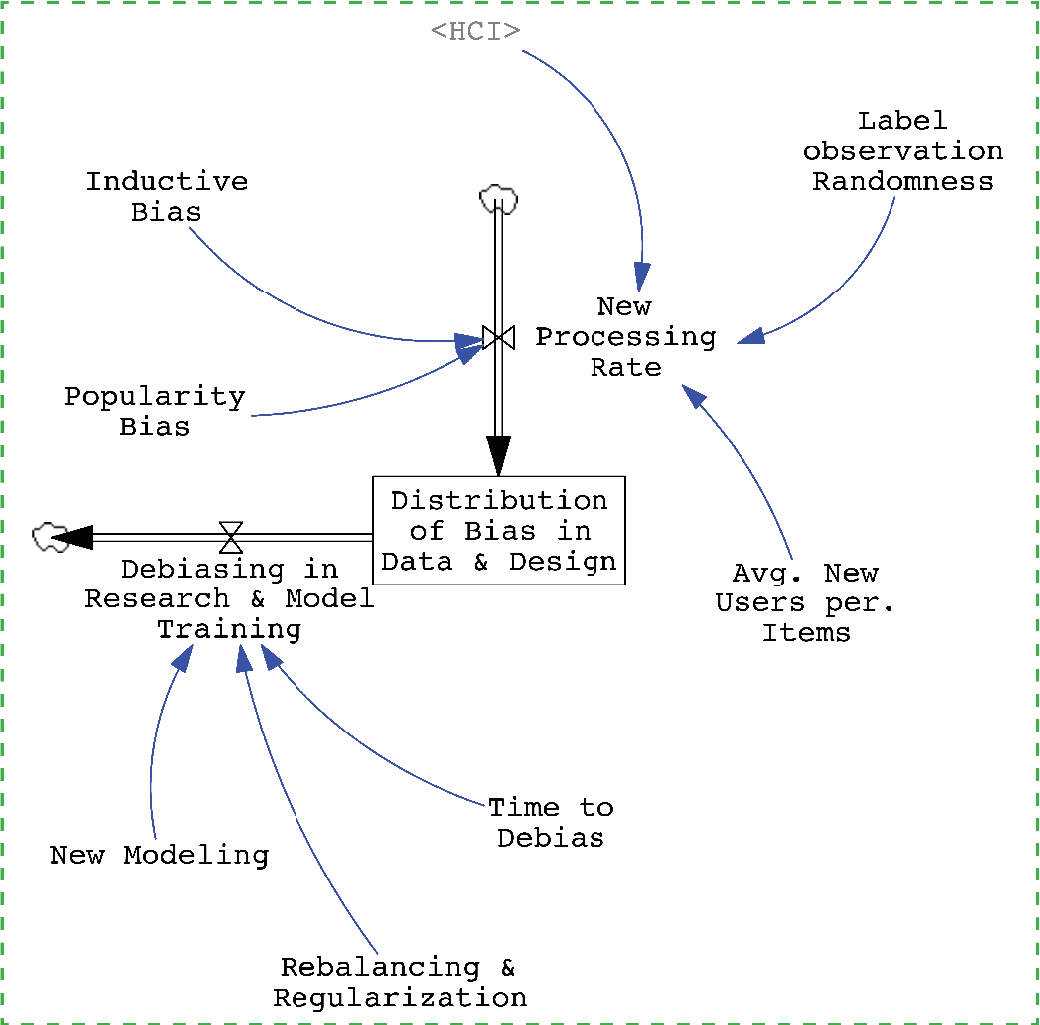}
    \caption{Bias Stock \& Flow}
    \label{fig:Model_left}
\end{figure}

\begin{figure}[!h]
    \centering
    \includegraphics[width=0.9\textwidth]{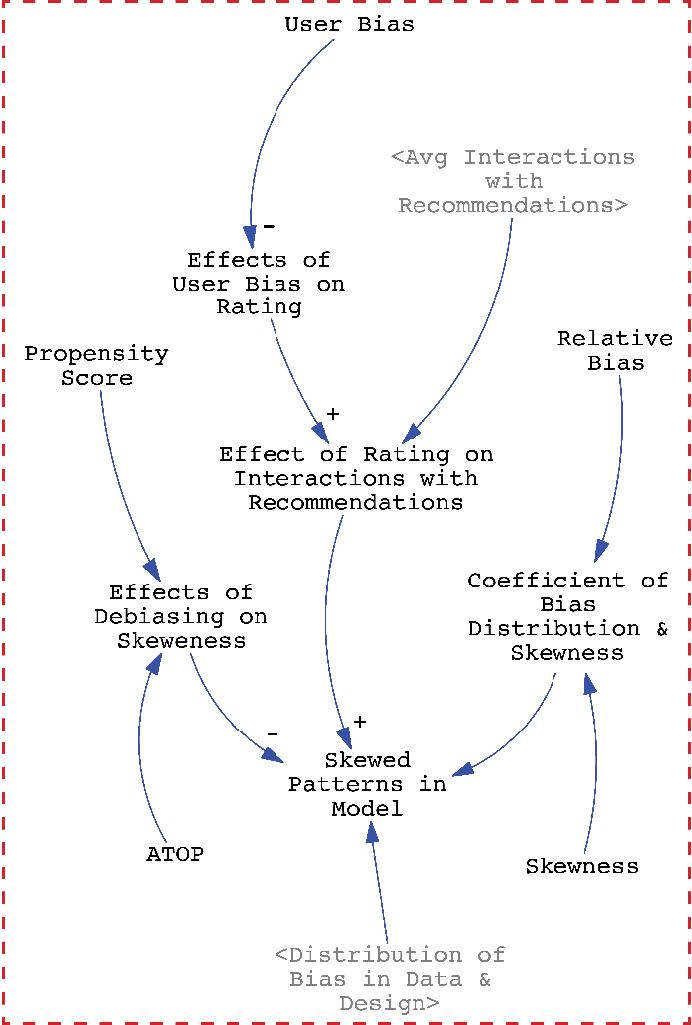}
    \caption{Factors Affecting Quality of Recommendations}
    \label{fig:Model_center}
\end{figure}

\begin{figure}[!h]
    \centering
    \includegraphics[width=\textwidth]{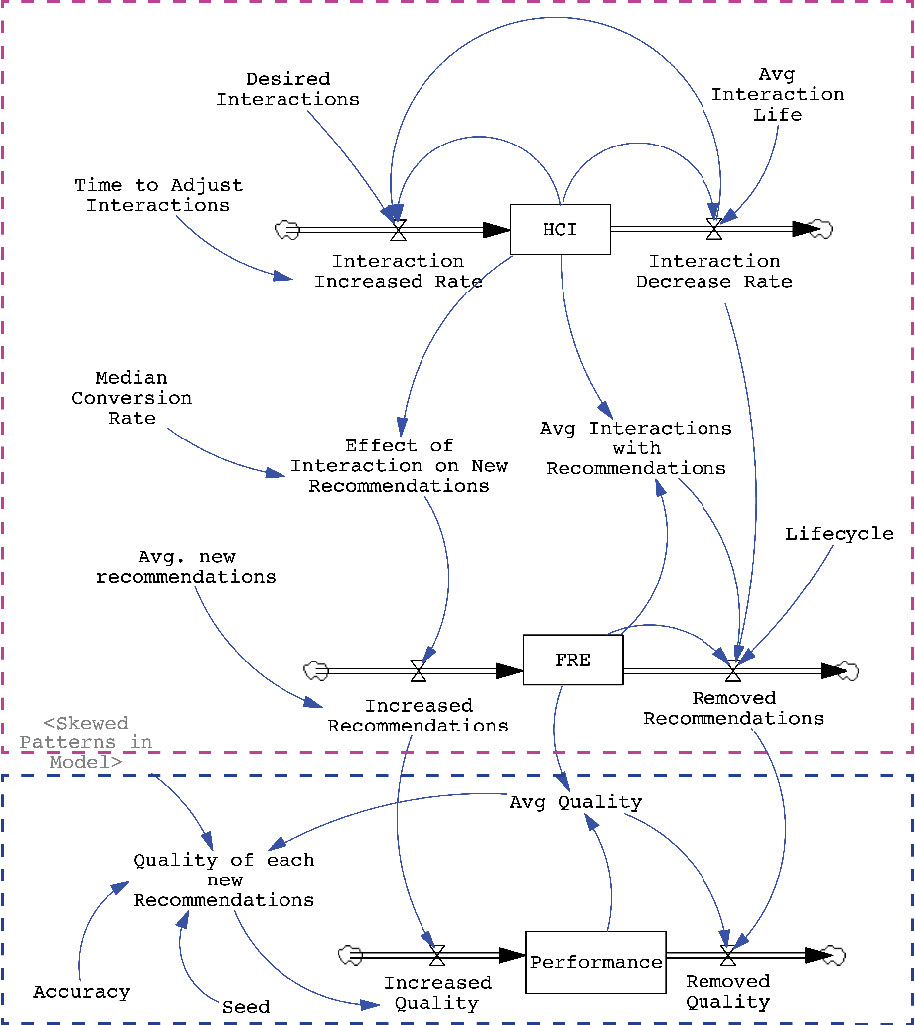}
    \caption{Recommender Engine, Interactions, and Performance Stock \& Flow}
    \label{fig:Model_right}
\end{figure}

\newpage

\section{Complete list of equations}\label{secA2}
The system dynamics methodology is reproducible through modeling with Vensim software using the equations provided:

\hfill \break
\noindent(01) Accuracy= 1

Units: Dmnl

\noindent(02) ATOP= 1 

Units: bias
    
\noindent(03) Avg Interaction Life= 6760

Units: Day
    
\noindent(04) Avg Interactions with Recommendations= FRE/HCI

Units: recommendations/interactions 
    
\noindent(05) Avg Quality=Performance/FRE

Units: quality/recommendations
    
\noindent(06) "Avg. new recommendations"=26000

Units: recommendations 
    
\noindent(07) "Avg. New Users per. Items"=1.74

Units: 1/Day
    
\noindent(08) "Coefficient of Bias Distribution \& Skewness"=Skewness/Relative Bias
    
Units: quality/bias]
    
\noindent(09) "Debiasing in Research \& Model Training"="Rebalancing \& Regularization"/(New Modeling*Time to Debias)

Units: bias/(interactions*Day)
    
\noindent(10) Desired Interactions=26000

Units: interactions
    
\noindent(11) "Distribution of Bias in Data \& Design"= INTEG (New Processing Rate-"Debiasing in Research \& Model Training",1)

Units: bias/interactions
    
\noindent(12) Effect of Interaction on New Recommendations=HCI*Median Conversion Rate

Units: 1/Day
    
\noindent(13) Effect of Rating on Interactions with Recommendations=Effects of User Bias on Rating/Avg Interactions with Recommendations

Units: interactions/(recommendations*bias)
 
\noindent(14) Effects of Debiasing on Skeweness=ATOP+Propensity Score

Units: bias

\noindent(15) Effects of User Bias on Rating=1/User Bias

Units: 1/bias

\noindent(16) FINAL TIME = 100 

Units: Day

The final time for the simulation. 

\noindent(17) FRE= INTEG ( Increased Recommendations-Removed Recommendations,5)

Units: recommendations

\noindent(18) HCI= INTEG (Interaction Increased Rate-Interaction Decrease Rate,10) 

Units: interactions

\noindent(19) Increased Quality=Quality of each new Recommendations*Increased Recommendations 

Units: quality/Day

\noindent(20) Increased Recommendations= Effect of Interaction on New Recommendations*"Avg. new recommendations" 

Units: recommendations/Day

\noindent(21) Inductive Bias= 1

Units: bias

\noindent(22) INITIAL TIME = 0 

Units: Day 

The initial time for the simulation. 
 
\noindent(23) Interaction Decrease Rate=HCI/Avg Interaction Life 

Units: interactions/Day 

\noindent(24) Interaction Increased Rate= 
MAX(0,(Desired Interactions-HCI)/Time to Adjust Interactions+Interaction Decrease Rate) 

Units: interactions/Day

\noindent(25) Label observation Randomness=1 

Units: Dmnl

\noindent(26) Lifecycle=180 

Units: Day

\noindent(27) Median Conversion Rate= 2.4 

Units: 1/(Day*interactions) 

\noindent(28) New Modeling=1 

Units: interactions 
 
\noindent(29) New Processing Rate= (Inductive Bias+Popularity Bias)*"Avg. New Users per. Items"/HCI*Label observation Randomness 

Units: bias/(interactions*Day)

\noindent(30) Performance= INTEG ( Increased Quality-Removed Quality,1) 

Units: quality

\noindent(31) Popularity Bias=1 

Units: bias

\noindent(32) Propensity Score=1 

Units: bias 
 
\noindent(33) Quality of each new Recommendations= 
 RANDOM NORMAL( 1 , 5 , Accuracy*Avg Quality , Skewed Patterns in Model , Seed ) 

Units: quality/recommendations

\noindent(34) "Rebalancing \& Regularization"= 0 

Units: bias

\noindent(35) Relative Bias=1 

Units: bias

\noindent(36) Removed Quality= Avg Quality*Removed Recommendations 

Units: quality/Day

\noindent(37) Removed Recommendations= 
FRE/Lifecycle+(Avg Interactions with Recommendations*Interaction Decrease Rate)

Units: recommendations/Day

\noindent(38) SAVEPER = TIME STEP 

Units: Day [0,?] 

The frequency with which output is stored.

\noindent(39) Seed=1 

Units: Dmnl 
 
\noindent(40) Skewed Patterns in Model= 
 (Effects of Debiasing on Skeweness*Effect of Rating on Interactions with Recommendations)* 
("Distribution of Bias in Data \& Design"*"Coefficient of Bias Distribution \& Skewness")

Units: quality/recommendations

\noindent(41) Skewness=1 

Units: quality

\noindent(42) TIME STEP = 0.0078125 

Units: Day [0,?] 

The time step for the simulation.

\noindent(43) Time to Adjust Interactions=6760 

Units: Day
 
\noindent(44) Time to Debias=1 

Units: Day

\noindent(45) User Bias=1 

Units: bias
%An appendix contains supplementary information that is not an essential part of the text itself but which may be helpful in providing a more comprehensive understanding of the research problem or it is information that is too cumbersome to be included in the body of the paper.

%%=============================================%%
%% For submissions to Nature Portfolio Journals %%
%% please use the heading ``Extended Data''.   %%
%%=============================================%%

%%=============================================================%%
%% Sample for another appendix section			       %%
%%=============================================================%%

%% \section{Example of another appendix section}\label{secA2}%
%% Appendices may be used for helpful, supporting or essential material that would otherwise 
%% clutter, break up or be distracting to the text. Appendices can consist of sections, figures, 
%% tables and equations etc.

\end{appendices}

%%===========================================================================================%%
%% If you are submitting to one of the Nature Portfolio journals, using the eJP submission   %%
%% system, please include the references within the manuscript file itself. You may do this  %%
%% by copying the reference list from your .bbl file, paste it into the main manuscript .tex %%
%% file, and delete the associated \verb+\bibliography+ commands.                            %%
%%===========================================================================================%%

\end{document}